\documentclass{sig-alternate-10pt}
\usepackage{amsmath}
\usepackage{graphicx}
\usepackage{color}
\usepackage{subfigure}
\usepackage{url}
\usepackage{xcolor}
\usepackage{booktabs}
\usepackage{amsfonts}
\usepackage{epsfig}
\usepackage{amscd}
\usepackage{amssymb}
\usepackage{graphicx}
\usepackage{bbding}
\usepackage{pifont}

\newcommand{\para}[1]{{\noindent\bf#1.~}}

\newcommand{\scloud}{{\sc SocialCloud}}
\title{SocialCloud: Using Social Networks for Building\\ Distributed Computing Services}
\numberofauthors{3}
\author{
\alignauthor
Abedelaziz Mohaisen\\
\affaddr{University of Minnesota}\\
\affaddr{Minneapolis, Minnesota 55455}
\and 
\alignauthor
Huy Tran\\
\affaddr{University of Minnesota}\\
\affaddr{Minneapolis, Minnesota 55455}
\and
\alignauthor
Abhishek Chandra\\
\affaddr{University of Minnesota}\\
\affaddr{Minneapolis, Minnesota 55455}
\and
\alignauthor
Yongdae Kim\\
\affaddr{University of Minnesota}\\
\affaddr{Minneapolis, Minnesota 55455}
}

\begin{document}
\maketitle

\begin{abstract}
In this paper we investigate a new computing paradigm, called SocialCloud, in which computing nodes are governed by social ties driven from a bootstrapping trust-possessing social graph. We investigate how this paradigm differs from existing computing paradigms, such as grid computing and the conventional cloud computing paradigms. We show that incentives to adopt this paradigm are intuitive and natural, and security and trust guarantees provided by it are solid. We propose metrics for measuring the utility and advantage of this computing paradigm, and using real-world social graphs and structures of social traces; we investigate the potential of this paradigm for ordinary users. We study several design options and trade-offs, such as scheduling algorithms, centralization, and straggler handling, and show how they affect the utility of the paradigm. Interestingly, we conclude that whereas graphs known in the literature for high trust properties do not serve distributed trusted computing algorithms, such as Sybil defenses---for their weak algorithmic properties, such graphs are good candidates for our paradigm for their self-load-balancing features.

\begin{keywords}
Distributed computing, Security, Trust, Social Computing, Performance. 
\end{keywords}

\end{abstract}

\section{Introduction}
\label{sec:intro}

Cloud computing is a new paradigm of computing that overcomes the restriction of conventional computing paradigms by enabling new technological and economical aspects, such as elasticity and pay-as-you-go---which free users from long-term commitments and obligation towards service providers. Cloud computing is beneficial for both consumers and cloud service providers. While it meets customers and users technological demands, the cloud computing paradigm is also a rich field of profit to cloud providers~\cite{Armbrust:2010aa}. 

For users, cloud computing overcomes several shortcomings as opposed to using conventional computing paradigms; where the used infrastructure and software are owned by the user. For example, cloud computing enables users of the cloud---who also can be providers of services---to virtually locate their contents closers to their consumers and reduce latency of serving such contents, a challenging issue in conventional computing settings. Also, considering the return on investment, cloud computing has its appealing economical benefits and incentives, which make it a desirable option to many users. These incentives can be seen in the long run as a reduced overall cost resulting from hardware and software liabilities and maintenance costs in alternative paradigms~\cite{Armbrust:2010aa}. As for providers, benefits are also economical in the absolute sense. 

The current conventional cloud computing paradigm has many benefits, despite posing several challenging issues that need to be addressed before wider adoption by many potential users~\cite{Dillon:2010aa}. Examples of these issues include the need for concrete and clear business model that outlines clearer service level agreements (SLA) and guarantees the rights of users~\cite{Iqbal:2010aa,Iqbal:2010ab,Clarke:2010aa}, the need for architectures that consider the variety of potential applications demanded by users, the need for programming models that consider the large scale of data in the cloud, and the need for new applications that benefit from the architectural and programming models in the cloud, among other issues. While many of these issues are being constantly addressed in ongoing research efforts; where several architectures~\cite{Dastjerdi:2010aa,Botts:2010aa,Wallis:2010aa}, programming models~\cite{Dean:2010aa, Condie:2010aa,Stonebraker:2010aa}, and applications~\cite{Pandey:2010aa,Wickremasinghe:2010aa,Hagiwara:2010aa,Wallis:2010aa,Buyya:2010aa,Iqbal:2010aa,Iqbal:2010ab} are proposed, security and data privacy are chief among other issues to be considered before this paradigm is widely accepted. Indeed, both outsider and insider threats to security and privacy of data in cloud systems are unlimited. Also, incentives do exist for cloud providers to make use of users' data residing in cloud for their own benefits, for the lack of regulations and enforcing policies.

In this paper, we oversee a new type of computing paradigm, called~\scloud, that enjoys parts of the merits provided by the conventional cloud. Imagine the scenario of a computing paradigm where users who collectively construct a pool of resources perform computational tasks on behalf of their social acquaintance. Our paradigm and model are similar in many aspects to the conventional grid-computing paradigm. It exhibits such similarities in that users can outsource their computational tasks to peers, complementarily to using friends for storage, which is extensively studied in literature. Our paradigm is, however, very unique in many aspects as well. Most importantly, our paradigm exploits the trust exhibited in social networks as a guarantee for the good behavior of other ``workers in the system''. Accordingly, the most important ingredient to our paradigm is the social bootstrapping graph, a graph that is used for recruiting workers for a social network. 

Indeed, social networks are very popular (c.f.~\textsection\ref{sec:sns}). This popularity of social networks has opened the door wide for investigating the potential of these networks for many applications. Problems that are unsolvable in the cyberspace are easily solvable using social networks, for that they possess both algorithmic properties---such as connectivity---and trust, which are used to reason about the behavior of honest users in the social network, and limit the misbehavior introduced by other malicious users supported by efficiency features. Most important to the context of our paradigm is the aggregate computational power of nodes in the social network. Indeed, beyond the nodes and social links, the social networks consist of users with computing machines that are idle for most of the time~\cite{barroso2007case}. Furthermore, owners of these computing machines are willing to share their computing resources for their friends, and for a different economical model than in the conventional cloud computing paradigm---fully altruistic one. This behavior makes our work share commonalities with an existing stream of work on creating computing services through volunteers~\cite{Weissman:2011:EED:1996014.1996019,chandra}. Our results hence highlight technical aspects of this direction and pose challenges for designs options when using social networks for recruiting such workers and enabling trust. 
\subsection{Contributions}
To this end, our contribution in this paper is mainly twofold:
\begin{itemize}

\item First, we investigate the potential of the social cloud computing paradigm by introducing a design that bootstraps from social graphs to construct distributing computing services. We advocate the merits of this paradigm over existing ones such as the grid computing paradigm. 
\item Second, we verify the potential of our paradigm using simulation set-up and real-world social graphs with varying social characteristics that reflect different, and possibly contradicting, trust models. Both graphs and the simulator are made public~\cite{mylink} to the community to make use of them, and improve by additional features.
\end{itemize}

\subsection{Organization} The organization of this paper is as follows. In \textsection\ref{sec:case} we argue for the case of our paradigm. In \textsection\ref{sec:prelim} we review the preliminaries of this work. In \textsection\ref{sec:design}, we introduce the main design, including an intensive discussion on the design options.  In \textsection\ref{sec:sim}, we describe our simulator used for verifying the performance aspects of our design. In \textsection\ref{sec:results} we introduce the main results and detailed analyses and discussion of the design options, their benefits, and limitations. In \textsection\ref{sec:related}, we summarize some of the related work, including work on using social networks for building trustworthy computing services.  In \textsection\ref{sec:discussion}, we we draw concluding remarks followed by future work and directions in \textsection\ref{sec:future}.

\section{The Case for \scloud}\label{sec:case}

In this paper, we look at the potential of using unstructured social graphs for building distributed computing systems. These systems are proposed with several anticipated benefits in mind. First, such systems would exploit locality of data based on the applications they are intended for, under the assumption that the data would be stored at multiple locations and shared among users represented in the social network---see~\textsection\ref{sec:usemodel} and~\cite{Weissman:2011:EED:1996014.1996019} for concrete examples of such applications. This is in fact not a far-fetched assumption. For example, consider a co-authorship social graph, like the one used in our experiments, where the \scloud~is proposed for deployment. In that scenario, data on which computations are to be performed is likely to be at multiple locations; on machines of research collaborators, co-authors, or previous co-authors. Even for some online social networks, the assumption and achieved benefits are not far-fetched as well, considering that friends would have similar interests, and likely to have contents replicated across different machines, which could be potentially of interest to use in our computing paradigm. Examples of such settings include photos taken at parties, videos---for image processing applications, among others.

The second advantage of this paradigm is its trustworthiness. In the recent literature, there has been a lot of interest in the distributed computing community for exploiting social networks to perform trustworthy computations. Examples of these literature works include exploiting social networks for cryptographic signing services~\cite{XuLP08}, Sybil defenses~\cite{YuKGF06,DanezisM09,YuGKX08}, and routing in many settings including the delay tolerant networks~\cite{1544047,1288113}. In all of these cases, along with the algorithmic property in these social networks, the built designs exploit the trust in social networks. The trust in these networks rationalizes the assumption of collaboration in these built system, and the tendency of nodes in the network to act according to the intended protocol with the theorized guarantees. Same as in all of these applications, \scloud~tries to exploit the trust aspect of the social network, and thus it is easy to reason about the behavior of nodes in this paradigm (c.f. \textsection\ref{sec:incentives}).

Related to trust exhibited in the social fabric utilized in our paradigm, the third advantage is that it is also easy to reason about the recruitment of workers. In this context, workers are nodes that are willing to perform computing tasks for other nodes (tasks outsourcers). This feature, when associated with the aforementioned trust, is quite advantageous when compared to the challenge of performing trustworthy computing on dedicated workers in the conventional grid-computing paradigm, where it is hard to recruit such workers. 

Finally, our design oversees an altruistic model of \scloud, where nodes participate in the system and do not expect in return. Further details on this model are in~\textsection\ref{sec:incentives}. 

\para{Grid Computing} While the \scloud~uses a similar paradigm to that of the grid computing paradigm---in the sense that both try to outsource computations and use high aggregate computational resources, the \scloud~is slightly different. In particular, in the \scloud, there is a pre-defined relationship between the task outsourcer and the computing worker, which does not exist in the grid-computing paradigm. We limit the computations to $1-$hop neighbors, which further improve trustworthiness of computations in our model.

\section{Assumptions and Settings}
\label{sec:prelim}
In this section, we review the preliminaries required for understanding the rest of this paper. In particular, we elaborate on the social networks, their popularity, and their potential for being used as bootstrapping tools for systems, services, and protocols. We describe the social network formulation at a high level, the economical aspect of our system, and finally, the attacker model.

\subsection{Systems on Social Networks}\label{sec:sns}
Social networks are so popular. Nine of the twenty most popular sites on the web are for social networking~\cite{ebizmba}. The top ten online social networking websites have more than 650 million of unique visitors per month in total. The most popular social network, Facebook~\cite{Facebook} alone serves 250 million unique visitors per month, with more than 96 unique visitors per second. Such popularity of social networks has motivated so many designs, protocols, and applications on top of social networks. Examples include routing~\cite{1544047,1288113,1281292,MartiGG04}, social gossip~\cite{1591824,1378547,1397752}, and Sybil defenses~\cite{YuKGF06} (c.f.~\textsection\ref{sec:related}). While they are different in the details of their operation, all of these designs and protocols weigh algorithmic properties (connectivity), trust, and collaboration in the underlying social networks, which are used for bootstrapping such systems. 

\subsection{Social Graphs---High Level Description}
In this paper we view the social network as an undirected and unweighted graph $G=(V,E)$, where $V=\{v_1,\dots,v_n\}$ is the set of vertexes, representing the set of nodes in the social graph, and correspond to users (or computing machines), and $E=\{e_{ij}\}$ (where $1\leq i\leq n$ and $1\leq j\leq n$) is the set of edges connecting those vertices---which implies that nodes associated with the social ties are willing to perform computations for each other. $|V|=n$ denotes the size of $G$ and $|E|=m$ denotes the number of edges in $G$.  In the rest of the paper, social network, network, and graph are used interchangeably to refer to both the physical computing network and the underlying bootstrapping social graph, and the meaning depends on the context. Also, we refer to computing entities associated with users in the social network as nodes.

\subsection{Economics of {SocialCloud}}\label{sec:incentives}
In our design we assume an altruistic model, which simplifies the behavior of users and arguments on the attacker model. In this altruistic model, users in the social network {\em donate} their {\em computing resources}---while not using them---to other users in the social network to use them for specific computational tasks. In return, the same users who donated their resources for others would anticipate others as well to perform their computations on behalf of them when needed.

One can further improve this model. Social networks are rich of trust characteristics that capture additional features, and can be used to rationalize this model in several ways. For example, trust in social networks, a well studied vein of research in this context~\cite{mohaisen11}, can be used to adjust this model so as users would bind their participation in computations to trust values that they assign to other users. In this work, in order to make use of and confirm this model, we limit outsourced computations at 1-hop. 

While we do not consider that in this paper, another model using interests and groups is worth mentioning for its popularity and potential as a future work. The incentives model can be further relaxed by enabling ``interest'' based model of computation where workers do computation to other nodes in the graph that only share some interest with them. This interest can be publicly identified by the membership of a node in a group. Investigating this model is left as a future work.

\subsection{Use Model and Applications}\label{sec:usemodel}
For our paradigm, we envision compute intensive applications, for which other systems have been developed in the past using different design principles, but lacking trust features; where trust is needed in such applications and provided by our paradigm. These systems include ones with resources provided by volunteers, as well as grid-like systems, like in Condor~\cite{litzkow1988condor}, MOON~\cite{Lin:2010:MMO:1851476.1851489}, Nebula~\cite{chandra,Weissman:2011:EED:1996014.1996019}, and SETI@Home~\cite{anderson2002seti}. 

Specific examples of applications built on top of these systems, that would as well fit to our use model, include blog analysis~\cite{Weissman:2011:EED:1996014.1996019}, web crawling and social-network applications (collaborative filtering, image processing, etc)~\cite{cardosa2011exploring}, scientific computing~\cite{wang2008scientific}, among others.

Notice that each of these applications requires certain levels of trust for which social ties are best suited as a trust bootstrapping and enabling tool. Especially, reasoning about the behavior of systems and expected outcomes (in a computing system in particular) would be well-served by this trust model. We notice that this social trust has been previously used as an enabler for privacy in file-sharing systems~\cite{isdal2010privacy}, anonymity in communications systems~\cite{Nagaraja07}, and collaboration in sybil defenses~\cite{lesniewski2008sybil,YuGKX08,mohaisen11}, among others. In this work, we use the same insight to propose a computing paradigm that relies on such trust and volunteered resources, in the form of shared computing time.
With that in mind, in the following section we elaborate on the attacker used in our system and trust models provided by our design, thus highlight its advantage and distancing our work from prior works in the literature.

%.what apps need trust in such environs, (ii) how do social ties enable that trust

\subsection{Attacker Model}
In this paper, as it is the case in many other systems built on top of social networks~\cite{YuGKX08,YuKGF06,sumup}, we assume that the attacker is restricted in many aspects. For example, the attacker has a limited capability of creating arbitrarily many edges between himself and other nodes in the social graph.  

While this restriction may contradict some recent results in the literature~\cite{1526784}---where it is shown that some legitimate users befriend random users in the social network who are potentially attackers, it can be relaxed to achieved the intended trust and attack model by considering an overlay of subset of friends of each users. This overlay expresses the trust value of the social graph well and eliminates the influence introduced by the attacker who infiltrated the social graph~\cite{mohaisen11}. For example, since each user decides on to which node among his adjacent nodes to outsource computations to, each user is aware of other users he knows well and those who are just social encounters that could be potential attackers.  Accordingly, the user himself decides whether to include a given node in his overlay or not, thus minimizing or eliminating harm and achieving the required trust and attack model. 

The description of the above attacker model might be at odds with the rest of the paper, especially that we use some online social networks that do not reflect characteristics of trust required in our paradigm. However, such networks, when used, are used for two reasons. First, to derive insight on the potential of such social networks, and others that share similar topological characteristics, for performing computational tasks according to the method devised in this paper. Second, we use them to illustrate that some of these social networks might be less effective than the trust-possessing social graphs, which we strongly advocate for our computing paradigm. 

\subsection{Trust in Grid Computing Systems} While there has been a lot of research on characterizing and improving trust in the conventional grid computing paradigm~\cite{azzedin2002evolving,azzedin2002towards, song2005trusted,kamvar2003eigentrust}---which is the closest paradigm to compare to ours, trust guarantees in such paradigm are less strict than what is expressed by social trust. For that, it is easy to see that some nodes in the grid computing paradigm may act maliciously by, for example, giving wrong computations, or refusing to collaborate; which is even easier to detect and tolerate, as opposed to acting maliciously~\cite{chakrabarti2007grid}. 

\section{The Design of SocialCloud}
\label{sec:design}
The main design of \scloud~is very simple, where complexities are hidden in design choices and options. In \scloud, the computing overlay is bootstrapped by the underlying social structure. Accordingly, nodes in the social graph act as workers to their adjacent nodes (i.e., nodes which are one hop away from the outsourcer of computations). An illustration of this design is depicted in Figure~\ref{fig:design}. In this design, nodes in the social graph, and those in the \scloud~overlay, use their neighbors to outsource computational tasks to them. For that purpose, they utilize local information to decide on the way they schedule the amount of computations they want each and every one of their neighbors to take care of. Accordingly, each node has a scheduler which she uses for deciding the proportion of tasks that a node wants to outsource to any given worker among her neighbors. Once a task is outsourced to the given worker, and assuming that both data and code for processing the task are transferred to the worker, the worker is left to decide how to schedule the task locally to compute it. Upon completion of a task, the worker sends back the computations result to the outsourcer.

\newpage
 \subsection{Design Options: Scheduling Entity}
In the \scloud, two schedulers are used. The first scheduler is used for determining the proportion of task outsourced to each worker and the second scheduler is used at each worker to determine how tasks outsourced by outsourcers are computed and in which order. While the latter scheduler can be easily implemented locally without impacting the system complexity, the decision used for whether to centralize or decentralize the former scheduler impacts the complexity and operation of the entire system. In the following, we elaborate on both design decisions, their characteristics, and compare them.
 
\subsubsection{Decentralized scheduler} 
\begin{figure}[htb]
\centering
\includegraphics[width=0.45\textwidth]{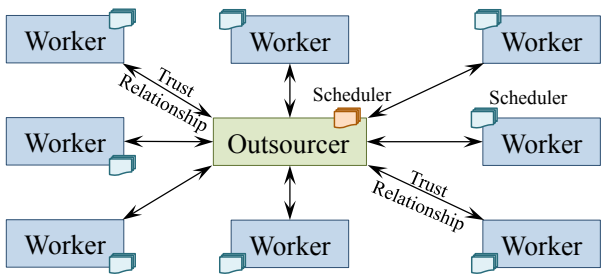} 
  \caption{A depiction of the main \scloud~paradigm as viewed by an outsourcer of computations. The different nodes in the social network act as workers for their friends, who act as potential jobs/tasks outsourcers. The links between social nodes are ideally governed by a strong trust relationship, which is the main source of trust for the constructed computing overlay. Both job outsourcers and workers have their own, and potentially different, schedulers.}\label{fig:design}
\end{figure}

In our paradigm, we limit selection of workers to 1-hop from the outsourcer. This makes it possible, and perhaps plausible, to incorporate scheduling of outsourcing tasks at the side of the outsourcer in a decentralized manner---thus each node takes care of scheduling its tasks. On the one hand, this could reduce the complexity of the design by eliminating the scheduling server in a centralized alternative. However, on the other hand, this could increase the complexity of the used protocols and the cost associated with them for exchanging {\em states}---such as availability of resources, online and offline time, among others. All of such states are exchanged between workers and outsourcers in our paradigm. These states are essential for building basic primitives in any distributed computing system to improve efficiency (see below for further details). An illustration of this design option is shown in Figure~\ref{fig:design}. In this scenario, each outsourcer, as well as worker, has its own separate scheduling component.

 \subsubsection{Centralized Scheduler} Despite the fact that nodes may only require their neighbors to perform the computational tasks on behalf of them and that may require only local information---which could be available to these nodes in advance, the use of a centralized scheduler might be necessitated to reduce communication overhead at the protocol level. 
 
 For example, in order to decide upon the best set of nodes to which to outsource computations, a node needs to know which of its neighbors are available, among other statistics. For that purpose, and given that the underlying communication network topology may not necessarily have the same proximity of the social network topology, the protocol among nodes needs to incur back and forth communication cost. 
 
 One possible solution to the problem is to use a centralized server that maintains states of the different nodes. Instead of communicating directly with neighbor nodes, an outsourcer would request the best set of candidates among its neighbors to the centralized scheduling server. In response, the server will produce a set of candidates, based on the locally stored states. Such candidates would typically be those that would have the most available resources to handle the outsourced computation task.

\begin{figure}[htb]
\centering
\includegraphics[width=0.45\textwidth]{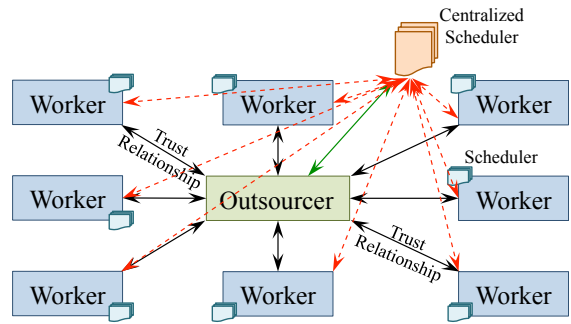} 
  \caption{The decentralized model of task scheduling in \scloud.}\label{fig:design2}
\end{figure}

An illustration of this design option is shown in Figure~\ref{fig:design2}. In this design, each node in \scloud~would periodically send states to a centralized server. When needed, an outsourcer node contacts the centralized server to return to it the best set of candidates for outsourcing computations, which the server would return based on the states of these candidates. Notice that only states are returned to the outsourcer, upon which the outsourcer would send tasks to these nodes on its own---Thus, the server involvement is limited to the control protocol.

The communication overhead of this design option to transfer states between a set of $d$ nodes is $2d$, where $d$ messages are required to deliver all nodes' states and $d$ messages are required to deliver states of all other nodes to each node in the set. On the other hand, $d(d-1)$ messages are required in the decentralized option (which requires pairwise communication of states update). When outsourcing of computations is possible among all nodes in the graph, this translates into $O(n)$ for the centralized versus $O(n^2)$ communication overhead for the decentralized option. To sum up, Table~\ref{tab:comp1} shows a comparison between both options.

\begin{table}[htb]
\begin{center}
\caption{A comparison between the centralized and decentralized scheduler options. Compared features are resistance to failure, communication overhead, required additional hardware, and required additional trust. F stands for failure, C stands for communication, H stands for hardware, and T stands for trust.}\label{tab:comp1} 
\begin{tabular}{ccccc}
\toprule
Option & F & C & H &T \\ 
\midrule
Centralized & \ding{54} & $O(n)$ & \ding{54}  & \ding{54}  \\
Decentralized & \ding{52}  & $O(n^2)$ &\ding{52}  & \ding{52} \\
\bottomrule
\end{tabular}
\end{center}
\end{table}
 
 \subsection{Tasks Scheduling Policy}\label{sec:taskpolicy}
 While the use of distributed or centralized scheduling entity resolves the issue of scheduling at the outsourcer side, two decisions remain unsolved: how much computation to outsource to each node (worker), and how much time a node among these workers should spend on a given task for a certain outsourcer. We handle these two issues separately. 
 
 As mentioned earlier, any off-the-shelf scheduling algorithm can be utilized to decide the right scheduling policy at the side of the outsourcer, which can be further improved by incorporating trust characterization models for weighted job scheduling~\cite{mohaisen11}. On the other hand, for workers scheduling, we consider several scheduling options as follows (notice that all of these policies are applied with respect to ``computing time''. This further requires estimating the time required for each task as a first step for using these policies). 
 \begin{itemize}
 \item \para{Round Robin (RR) Scheduling Policy} This is the simplest policy to implement, in which a worker spends an equal share of time on each outsourced task in a round robin fashion among all tasks he has.
 \item \para{Shortest First (SF)  Scheduling Policy} The worker performs shortest task first.  
 \item \para{Longest First (LF)  Scheduling Policy} The worker performs longest task first.
 \end{itemize}
Notice that we omit a lot of details about the underlying computing infrastructure, and abstract such infrastructure to ``time sharing machines'', which further simplifies much of the analysis in this work. In the results, we experiment with the three scheduling policies. %Notice that these policies are by no means enforced, though we use them to get insight on the operation of~\scloud.
 \subsection{Handling Outliers}\label{sec:outlier}
 The main performance criterion used for evaluating \scloud~is the time required to finish computing tasks for all nodes with tasks in the system. Accordingly, an outlier (also called a computing straggler) is a node with computational tasks that take a long time to finish, thus increasing the overall time to finish and decreasing the performance of the overall system. Detecting outliers in our system is simple: since the total time is given in advance, outliers are nodes with computing tasks that have longer time to finish when other nodes participating in the same outsourced computation are idle. 
 
Our method for handling outliers is simple too: when an outlier is detected, we outsource the remaining part of computations on all idle nodes neighboring the original outsourcer. For that, we use the same scheduling policy used by the outsourcer when she first outsourced this task. In the simulation part, we consider both scenarios of handled and unhandled outliers, and observe how they affect the performance of the system.
 
\subsection{Deciding Workers Based on Resources}
In real-world deployment of a system like \scloud, we expect heterogeneity of resources, such as bandwidth, storage, and computing power, in workers. This heterogeneity would result in different results and utilization statistics of a system like~\scloud, depending on which nodes are used for what tasks. 

While our work does not address this issue, and leaves it as a future work (c.f.~\textsection\ref{sec:limits} and~\textsection\ref{sec:discussion}). We further believe that simple decisions can be made in this regard so as to meet the design goals and achieve the good performance. For example, we expect that nodes would select workers among their social neighbors that have resources and link capacities exceeding a threshold, thus meeting an expected performance.

\section{Simulator of \scloud}
\label{sec:sim}
\begin{figure}[htb]
\centering
\includegraphics[width=.49 \textwidth]{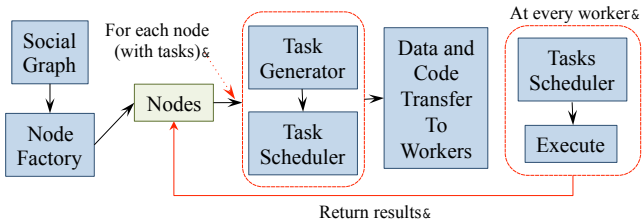}
  \caption{The flow diagram of \scloud: social graph is used for bootstrapping the computing service and recruit workers, nodes are responsible for scheduling their tasks by determining the amount of work each of its neighbors would process, and each worker (node) uses its local scheduler to determine how much time is allowed for each sub-task by its neighbors.}\label{fig:flow}
\end{figure}

To demonstrate the potential of~\scloud~as a computing paradigm, we implement a batch-based simulator~\cite{mylink} that considers a variety of scheduling algorithms, an outlier handling mechanism, job generation handling, and failure simulation. A flow diagram of the simulator is in Figure~\ref{fig:flow}. 

The flow of the simulator, which represents the flow of the system, is depicted in Figure~\ref{fig:flow}. First, the node factory uses the bootstrapping social graph to create nodes and their workers. Each node then decides on whether she has a task or not, and if she has a task she schedule the task according to her scheduling algorithm. If needed, each node then transfers code on which computations are to be performed to the worker along with the splits of the data for these codes to run on. Each worker then performs the computation according to the scheduling algorithm of the worker and returns the results of the computations to the outsourcer. 

\para{Timing} In \scloud, we use {\em virtual time} to simulate computations and resources sharing. We scale down the simulated time by 3 orders of magnitude of that in reality. This is, for every second worth of computations in real-world, we use one millisecond in the simulation environment. Thus, units of times in the rest of this paper are in virtual seconds.

\section{Results and Analysis}
\label{sec:results}
In this section, in order to derive insight on the potential of \scloud, we experiment with the simulator described above. Before getting into the details of the experiments, we describe the data and evaluation metric used in this section.

\subsection{Evaluation Metric}\label{sec:metric}
To demonstrate the potential of operating \scloud, we use the ``normalized finishing time'' of a task outsourced by a user to other nodes in the \scloud~as the performance metric. We consider the same metric over the different graphs used in the simulation. To demonstrate the performance for the population of all nodes that have tasks to be computed in the system, we use the empirical CDF (commutative distribution function) as an aggregate measure. For a random variable $X$, the CDF is defined as $F_X(x)=P_r(X\leq x)$. In our experiments, the CDF measures the fraction (or percent) of nodes that finish their tasks before a point in time $x$, as part of the overall number of tasks. We define $x$ as the factors of time of normal operation per dedicated machines, if they were to be used instead of outsourcing computations. This is, suppose that the overall time of a task is $T_{tot}$ and the time it takes to compute the subtask by the slowest worker is $T_{last}$, then $x$ for that node is defined as $T_{last}/T_{tot}$.

\subsection{Tasks Generation and Weights}\label{sec:taskgen}
Also for demonstrating the operation of our simulator, and the trade-off that such operation provides, we consider two different approaches for the tasks generated by each user. The size of each generated task is measured by virtual units of time, and for our demonstration we use two different scenarios:
\begin{itemize}
\item \para{Constant task weight} each outsourcer generates tasks with an equal size. These tasks are divided into equal shares and distributed among different workers in the computing system. The size of each task is $\bar{T}$. 
\item \para{Variable task weight} each outsourcer has a different task size. We model the size of tasks as a uniformly distributed random variable in the range of $[\bar{T}-\ell, \bar{T}+\ell]$ for some $\bar{T}>\ell$. Each worker receives an equal share of the task from the outsourcer.
\end{itemize}
\subsection{Deciding Tasks Outsourcers}
Not all nodes in the system are likely to have tasks to outsource for computation at the same time. Accordingly, we denote the fraction of nodes that have tasks to compute by $p$, where $0<p<1$. In our experiments we use $p$ from $0.1$ to $0.5$ with increments of $0.1$. We further consider that each node in the network has a task to compute with probability $p$, and has no task with probability $1-p$---thus, whether a node has a task to distribute among its neighbors and compute or not follows a binomial distribution with a parameter $p$. Once a node is determined to be among nodes with tasks at the current round of run of the simulator, we fix the task length. For tasks length, we use both scenarios mentioned in \textsection\ref{sec:taskgen}; with fixed or constant and variable tasks weights.

\subsection{Social Graphs}
To derive insight on the potential of \scloud, we run our simulator on several social graphs with different size and density, as shown in Table \ref{tab:datasets}. The graphs used in these experiments represent three co-authorship social structures (DBLP, Physics 1, and Physics 2), one voting network (of Wiki-vote for wikipedia administrators election), and one friendship network (of the consumer review website, Epinion). All of these graphs are made undirected, if they are not already, which rationalizes their use in our system. Notice the varying density of these graphs, which also reflects on varying topological characteristics. Also, notice the nature of these social graphs, where they are built in different social contexts and possess varying qualities of trust~\cite{mohaisen11}.

Next, we present the main results and findings of our design when operated on these graphs.

\begin{table}[htb]
\begin{center}
\caption{Social graphs used in our experiments.}\label{tab:datasets}
\begin{tabular}{llll}
\toprule
Dataset & \# nodes & \# edges & Description \\
\midrule
DBLP& 614981 & 1155148 & CS Co-authorship\\
Epinion & 75877 & 405739 & Friendship network\\
Physics 2 & 11204 &117649 &Co-authorship\\
Wiki-vote & 7066 & 100736 &Voting network\\
Physics 1 & 4158 & 13428 & Co-authorship\\
\bottomrule
\end{tabular}\vspace{-5mm}
\end{center}
\end{table}

\subsection{Main Results}
In this section we demonstrate our paradigm and discuss the main results of this work. Due to the lack of space, we delegate additional results to the technical report in~\cite{thetr}. For all measurements, our metric of performance and comparison is the normalized time to finish metric, explained in section~\ref{sec:metric}.
\begin{figure*}[htb]
\begin{center}
	\subfigure[Physics 1.] {\includegraphics[width=.32\textwidth]{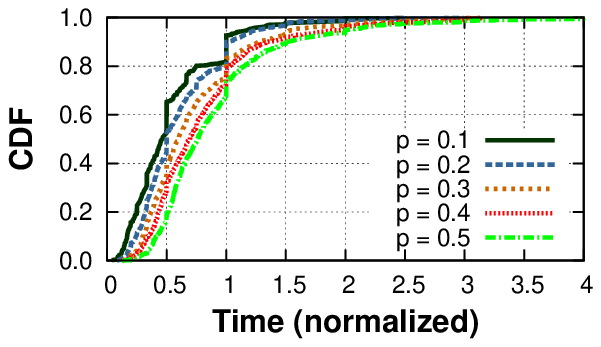}\label{pphy1}}
	\subfigure[Physics 2.]{\includegraphics[width=.32\textwidth]{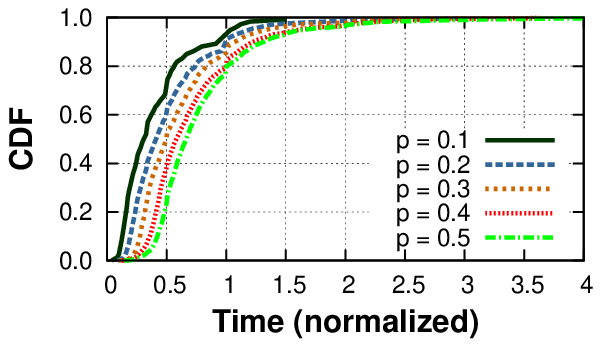}\label{pphy2}}
	\subfigure[DBLP.]{\includegraphics[width=.32\textwidth]{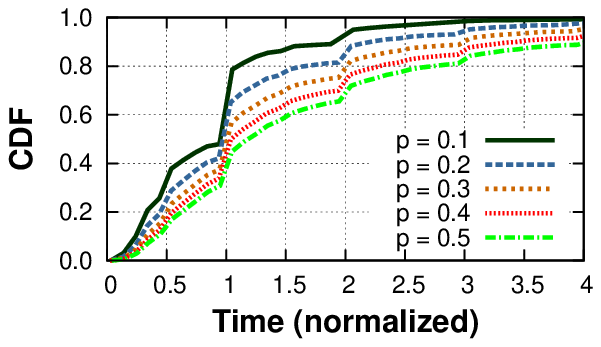}\label{pdblp}}
	\subfigure[Epinion.]{\includegraphics[width=.32\textwidth]{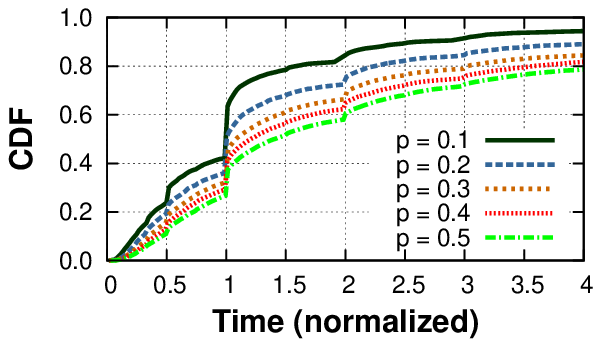}\label{pep}}
	\subfigure[Wiki-vote.]{\includegraphics[width=.32\textwidth]{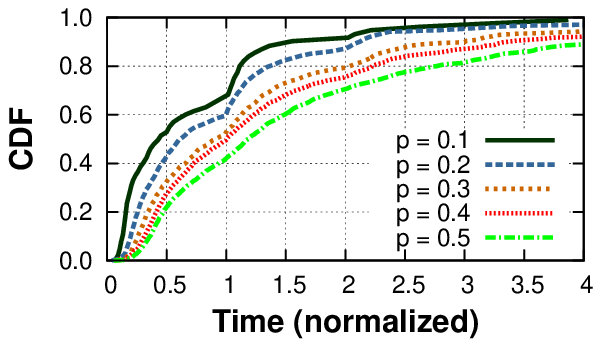}\label{pwiki}}
\end{center}
\caption{The normalized time it takes to perform outsourced computations in \scloud. Different graphs with different social characteristics have different performance results, where those with well-defined social structures have self-load-balancing features, in general. These measurements are taken with round-robin scheduling algorithm that uses the outlier handling policy in \textsection\ref{sec:outlier} for a fixed task size (of 1000 simulation time units).}\label{fig:p}
\end{figure*}
\subsubsection{Performance When Varying the Number of Outsourcers} In the first experiment, we run our \scloud~simulator on the different social graphs discussed earlier to measure the evaluation metric when the number of the outsourcers of tasks increases. We consider $p=0.1$ to $0.5$ with increments of $0.1$ at each time. The results of this experiment are in Figure~\ref{fig:p}. On the results of this experiment we make several observations. 

First, we observe the potential of~\scloud, even when the number of outsourcers of computations in the social network is as high as $50\%$ of the total number of nodes, which translates into a small normalized time to finish even in the worst performing social graphs (about $60\%$ of all nodes with tasks would finish in 2 normalized time units). However, this advantage varies for different graphs: we observe that sparse graphs, like co-authorship graphs, generally outperform other graphs used in the experiments (by observing the tendency in the performance in figures~\ref{pphy1} through~\ref{pdblp} versus figures~\ref{pep} and~\ref{pwiki}). In the aforementioned graphs, for example, we see that when $10\%$ of nodes in each case is used, and by fixing $x$, the normalized time, to $1$, the difference of performance is about $30\%$. This difference of performance is observed between the Physics co-authorship graphs---where $95\%$ of nodes finish their computations---and the Epinion graph---where only about $65\%$ of nodes finish their computations. 

\begin{figure*}[htb]
\centering
\subfigure[Handled outliers ($p=0.1$)]{\includegraphics[width=0.32\textwidth]{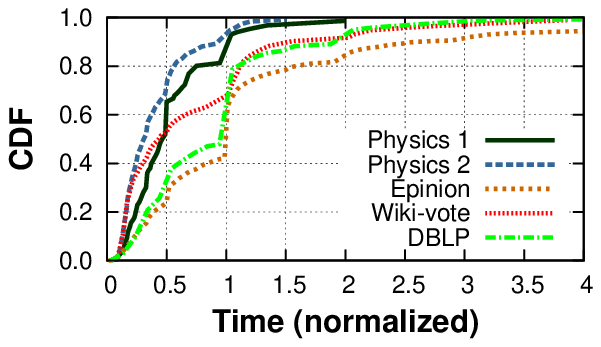}\label{allh01}}
\subfigure[Handled outliers ($p=0.3$)]{\includegraphics[width=0.32\textwidth]{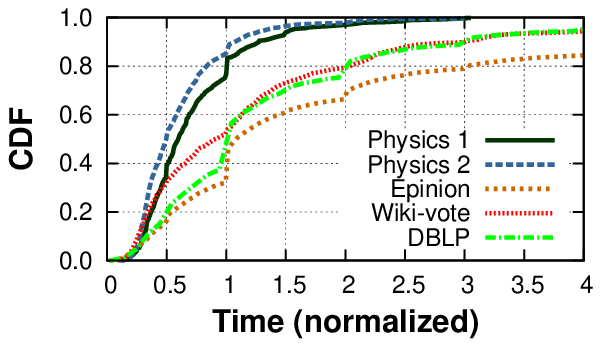}\label{allh03}}
\subfigure[Handled outliers ($p=0.5$)]{\includegraphics[width=0.32\textwidth]{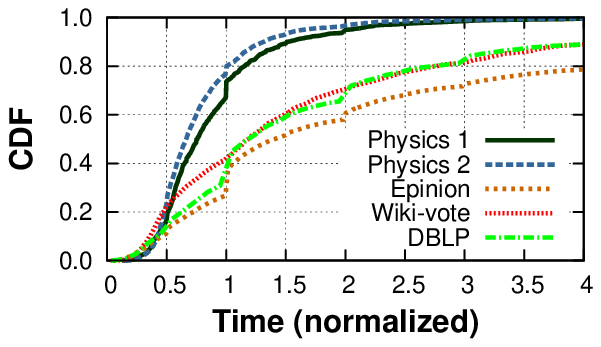}\label{allh05}}
\subfigure[Unandled outliers  ($p=0.1$)]{\includegraphics[width=0.32\textwidth]{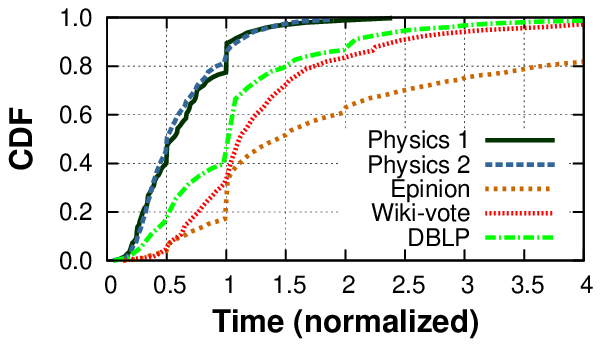}\label{alluh01}}
\subfigure[Unhandled outliers  ($p=0.3$)]{\includegraphics[width=0.32\textwidth]{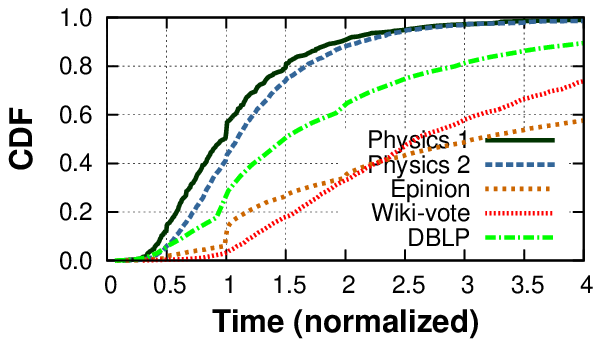}\label{alluh03}}
\subfigure[Unhandled outliers  ($p=0.5$)]{\includegraphics[width=0.32\textwidth]{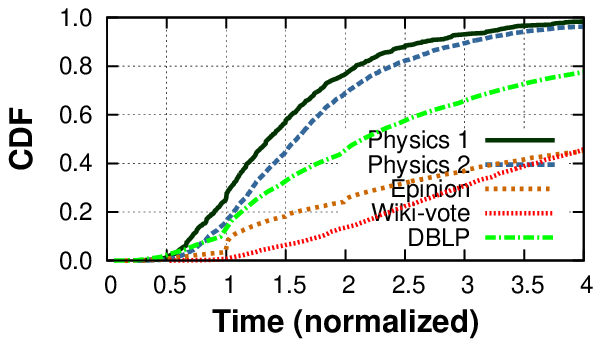}\label{allhu05}}
  \caption{The performance of \scloud~on the different social graphs used for our experiments, demonstrating the inherent differences in the different social graphs. Both figures use $p=0.3$ and the round robin scheduling algorithm.}\label{fig:all03B}
\end{figure*}

Second, we observe that the impact of $p$, the fraction of nodes with tasks in the system, would depend on the graph rather than $p$ alone. For example, in Figure~\ref{pphy1}, we observe that moving from $p=0.1$ to $p=0.5$ (when $x=1$) leads to a decrease in the fraction of nodes that finish their computations from $95\%$ to about $75\%$. On the other hand, for the same settings, this would lead to a decrease from about $80\%$ to $40\%$, a decrease from about $65\%$ to $30\%$, and a decrease from $70\%$ to $30\%$ in DBLP, Epinion, and Wiki-vote, respectively. This suggests that the decreases in the performance are due to an inherit property of each graph. The inherit property of each graph and how it affects the performance of~\scloud~is further illustrated in Figure~\ref{fig:all03B}. Interestingly, we find that even if DBLP is almost two orders of magnitude the size of Wiki-vote, for example, it outperforms Wiki-vote when not using outlier handling, and gives almost the same performance when using outliers handling.

\begin{figure*}[htb]
\begin{center}
	\subfigure[Physics 1 ($p=0.1$).] {\includegraphics[width= 0.24\textwidth]{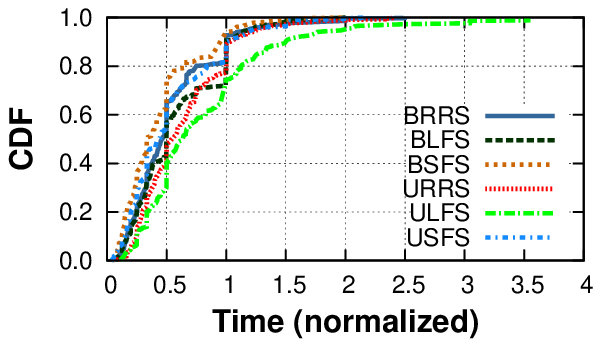}\label{ppp1}}
	\subfigure[Physics 2 ($p=0.1$).]{\includegraphics[width= 0.24\textwidth]{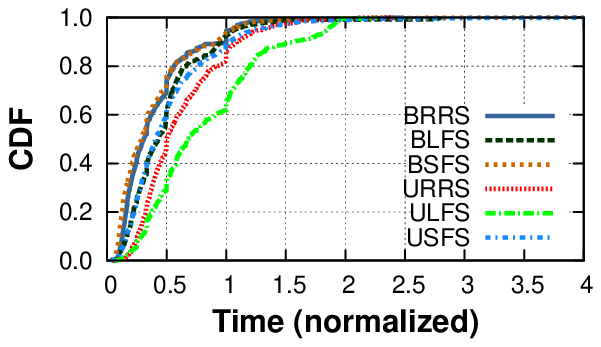}\label{subfig:phy2}}
	\subfigure[DBLP	($p=0.1$).]{\includegraphics[width= 0.24\textwidth]{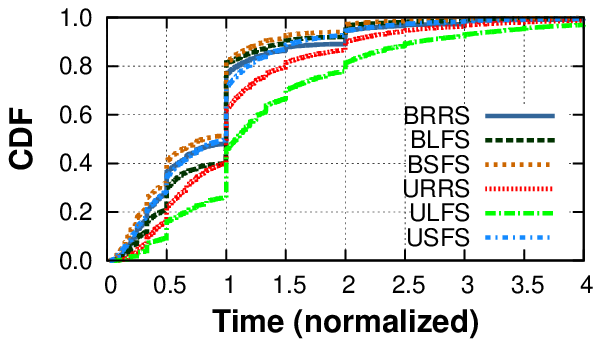}\label{subfig:phy3}}
	\subfigure[Epinion	($p=0.1$).]{\includegraphics[width= 0.24\textwidth]{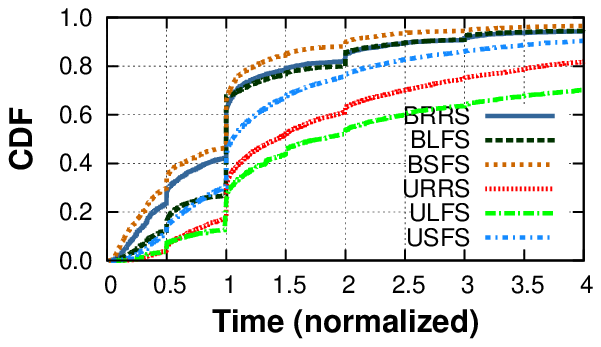}\label{subfig:phy3}}
	\subfigure[Wiki-vote	($p=0.1$).]{\includegraphics[width= 0.24\textwidth]{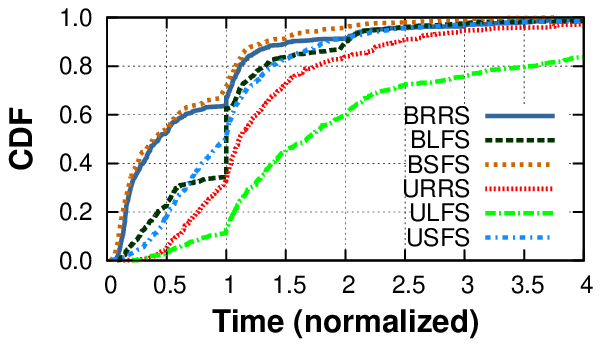}\label{subfig:phy3}}	
	\subfigure[Physics 1	($p=0.3$).] {\includegraphics[width=0.24\textwidth]{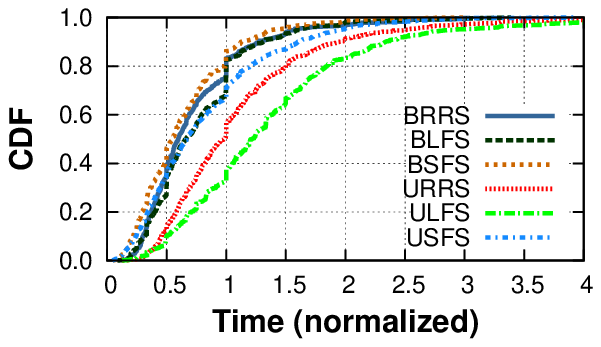}\label{subfig:phy1}}
	\subfigure[Physics 2	($p=0.3$).]{\includegraphics[width= 0.24\textwidth]{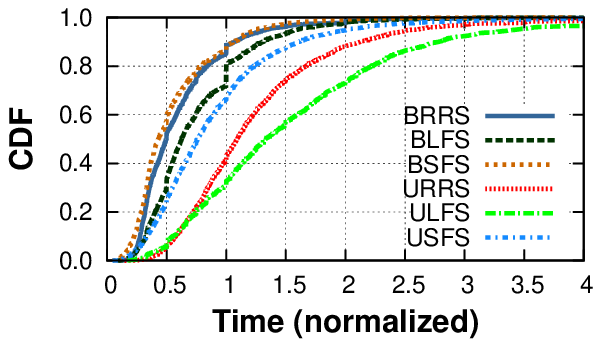}\label{subfig:phy2}}
	\subfigure[DBLP	($p=0.3$).]{\includegraphics[width= 0.24\textwidth]{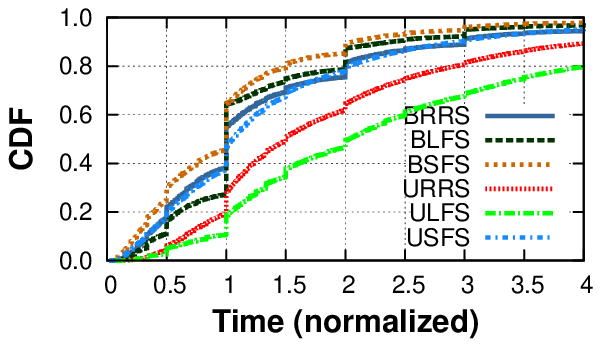}\label{subfig:phy3}}
	\subfigure[Epinion	($p=0.3$).]{\includegraphics[width= 0.24\textwidth]{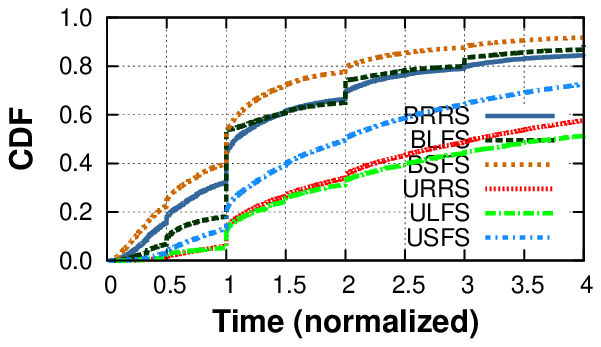}\label{subfig:phy3}}
	\subfigure[Wiki-vote	($p=0.3$).]{\includegraphics[width= 0.24\textwidth]{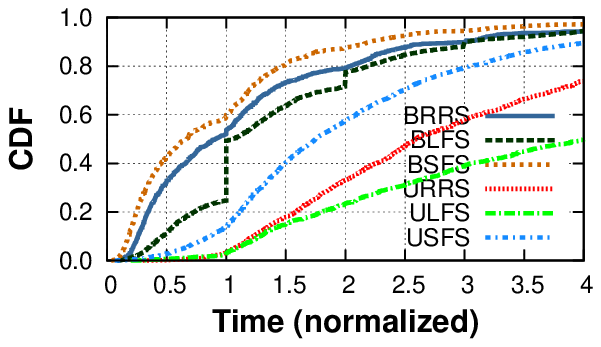}\label{subfig:phy3}}
	\subfigure[Physics 1	($p=0.5$).] {\includegraphics[width=0.24\textwidth]{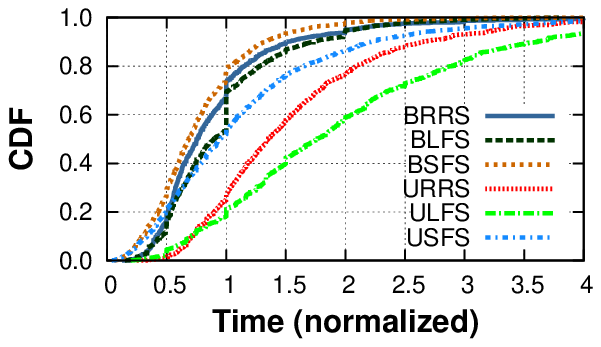}\label{subfig:phy1}}
	\subfigure[Physics 2	($p=0.5$).]{\includegraphics[width= 0.24\textwidth]{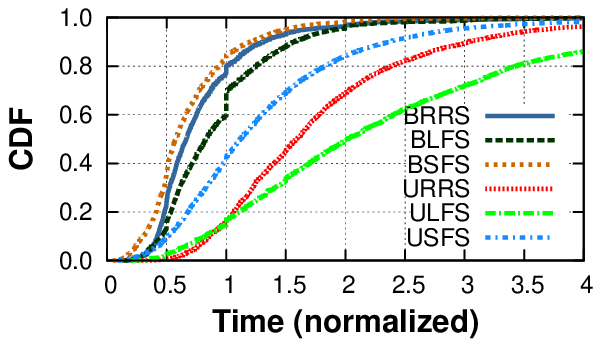}\label{subfig:phy2}}
	\subfigure[DBLP	($p=0.5$).]{\includegraphics[width= 0.24\textwidth]{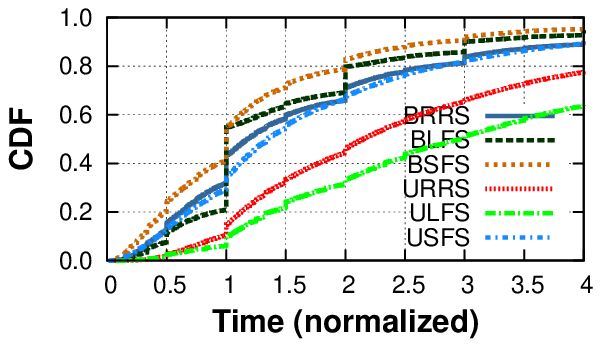}\label{subfig:phy3}}
	\subfigure[Epinion	($p=0.5$).]{\includegraphics[width= 0.24\textwidth]{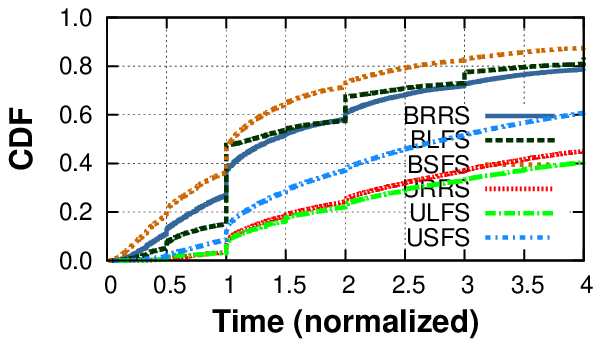}\label{subfig:phy3}}
	\subfigure[Wiki-vote	($p=0.5$).]{\includegraphics[width= 0.24\textwidth]{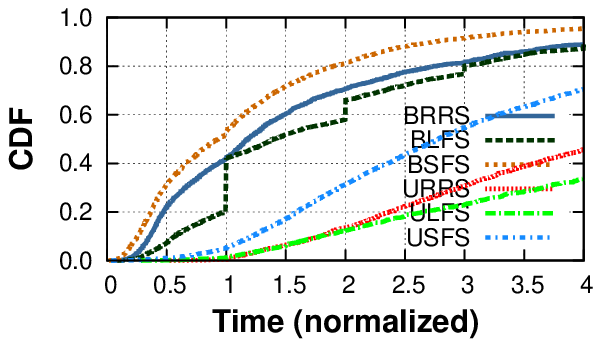}\label{ppp10}}
\end{center}
\caption{The normalized time it takes to perform outsourced computations in \scloud~for different scheduling policies. Naming convention: U stands for unhandled outlier and B stands for handled outliers (Balanced). RRS, SFS, and LFS stand for round-robin, shortest first, and longest first scheduling. We fix the job size among all outsourcers.}\label{fig:policy}
\end{figure*}

\begin{figure*}[htb]
\begin{center}
	\subfigure[Physics 1 ($p=0.1$).] {\includegraphics[width= 0.24\textwidth]{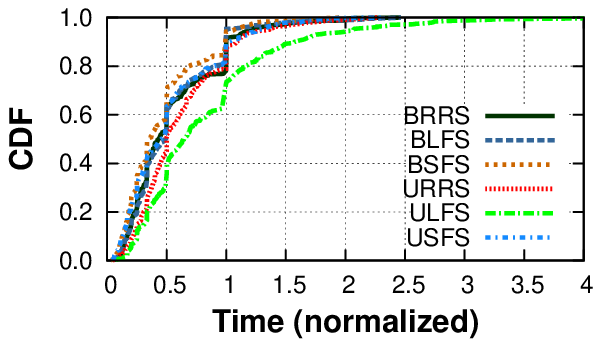}\label{ppp1v}}
	\subfigure[Physics 2 ($p=0.1$).]{\includegraphics[width= 0.24\textwidth]{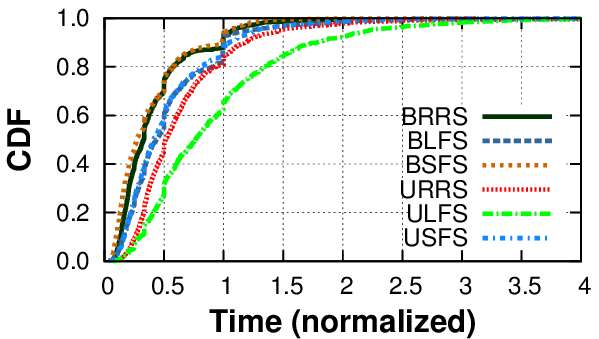}\label{subfig:phy2v}}
	\subfigure[DBLP	($p=0.1$).]{\includegraphics[width= 0.24\textwidth]{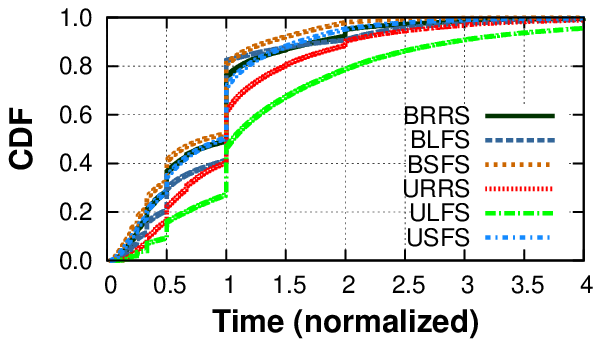}\label{subfig:phy3v}}
	\subfigure[Epinion	($p=0.1$).]{\includegraphics[width= 0.24\textwidth]{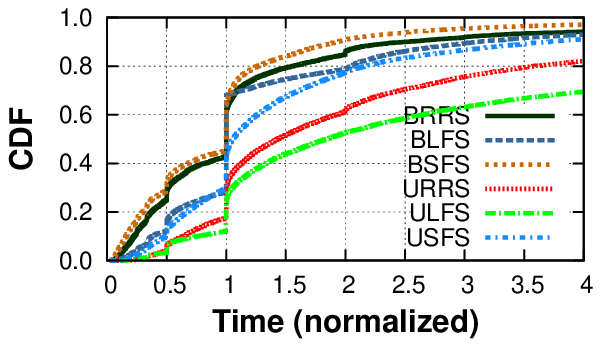}\label{subfig:phy3v}}
	\subfigure[Wiki-vote	($p=0.1$).]{\includegraphics[width= 0.24\textwidth]{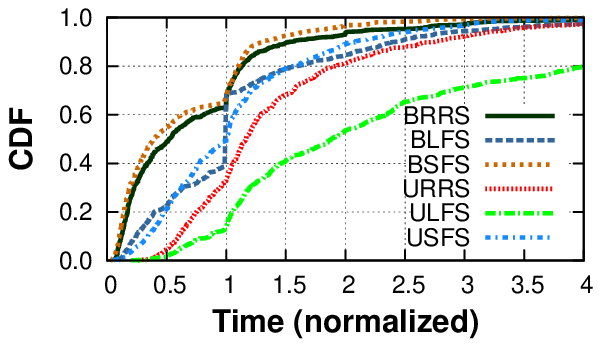}\label{subfig:phy3v}}	
	\subfigure[Physics 1	($p=0.3$).] {\includegraphics[width=0.24\textwidth]{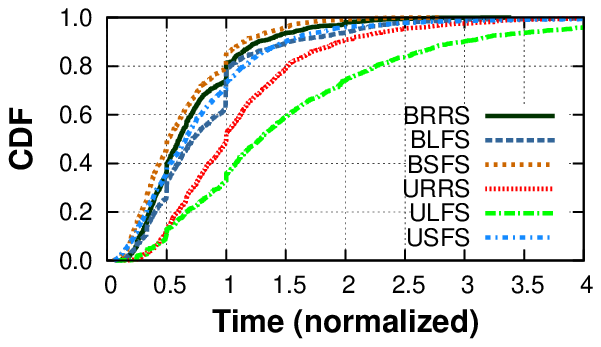}\label{subfig:phy1v}}
	\subfigure[Physics 2	($p=0.3$).]{\includegraphics[width= 0.24\textwidth]{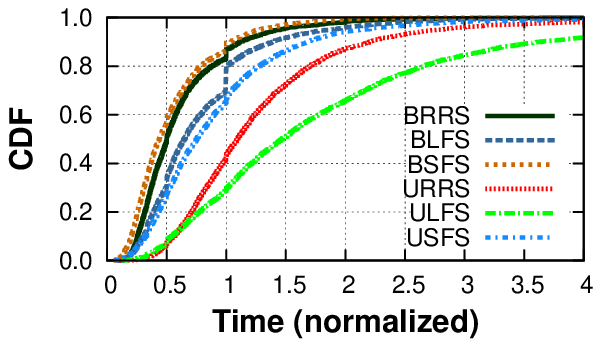}\label{subfig:phy2v}}
	\subfigure[DBLP	($p=0.3$).]{\includegraphics[width= 0.24\textwidth]{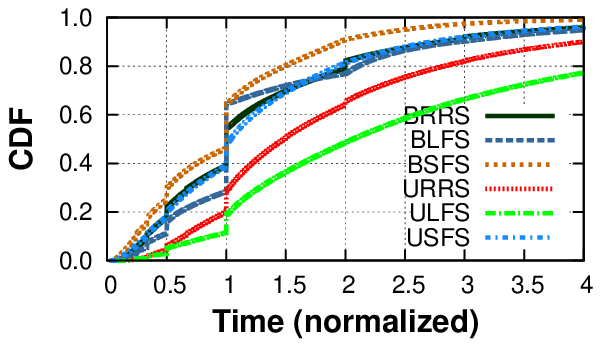}\label{subfig:phy3v}}
	\subfigure[Epinion	($p=0.3$).]{\includegraphics[width= 0.24\textwidth]{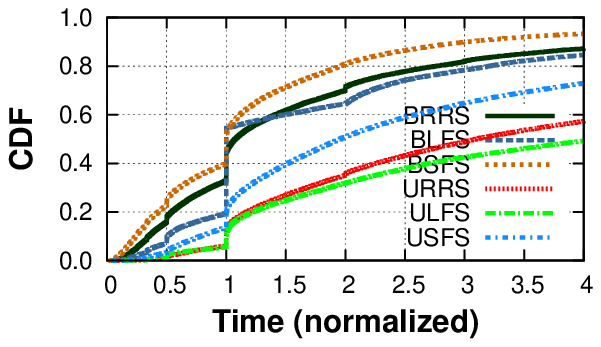}\label{subfig:phy3v}}
	\subfigure[Wiki-vote	($p=0.3$).]{\includegraphics[width= 0.24\textwidth]{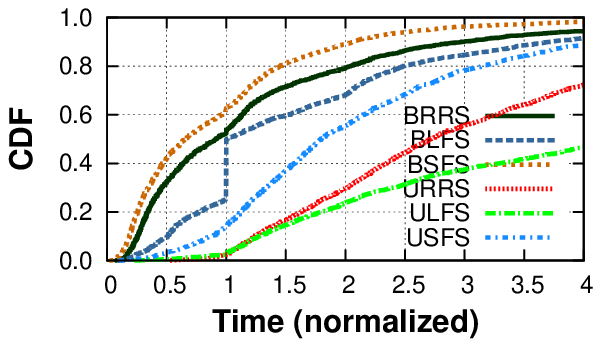}\label{subfig:phy3v}}
	\subfigure[Physics 1	($p=0.5$).] {\includegraphics[width=0.24\textwidth]{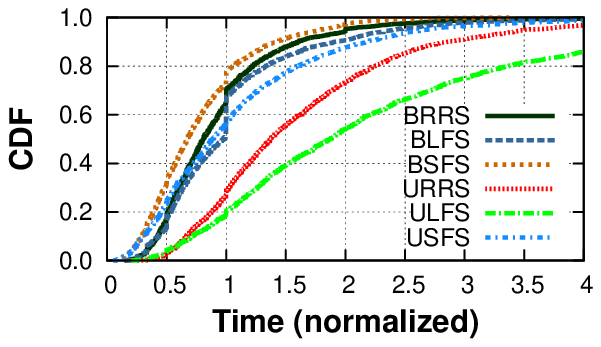}\label{subfig:phy1v}}
	\subfigure[Physics 2	($p=0.5$).]{\includegraphics[width= 0.24\textwidth]{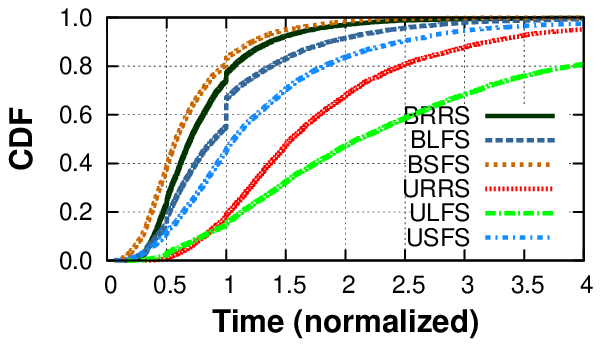}\label{subfig:phy2v}}
	\subfigure[DBLP	($p=0.5$).]{\includegraphics[width= 0.24\textwidth]{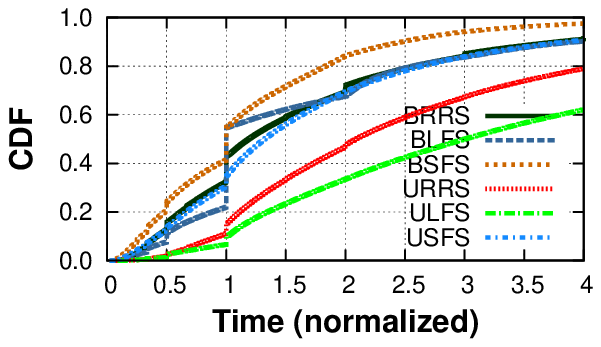}\label{subfig:phy3v}}
	\subfigure[Epinion	($p=0.5$).]{\includegraphics[width= 0.24\textwidth]{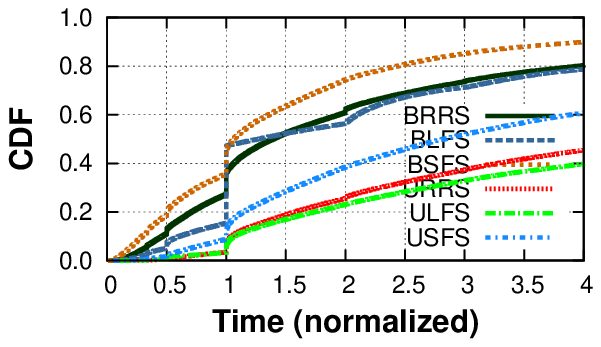}\label{subfig:phy3v}}
	\subfigure[Wiki-vote	($p=0.5$).]{\includegraphics[width= 0.24\textwidth]{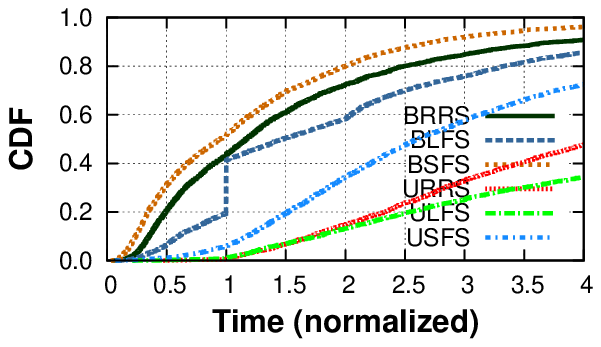}\label{ppp10v}}
\end{center}\vspace{-2mm}
\caption{The normalized time it takes to perform outsourced computations in \scloud~for different scheduling policies. Naming convention: U stands for unhandled outlier and B stands for handled outliers (Balanced). RRS, SFS, and LFS stand for round-robin, shortest first, and longest first scheduling. We sets jobs with variable lengths as described above.}\label{fig:policyX}
\end{figure*}

\begin{figure*}[htb]
\begin{center}
	\subfigure[Physics 1	($p=0.3$).] {\includegraphics[width=0.32\textwidth]{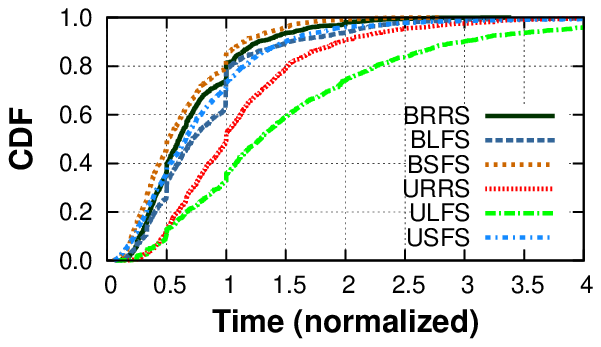}\label{subfig:phy1}}
	\subfigure[Physics 2	($p=0.3$).]{\includegraphics[width= 0.32\textwidth]{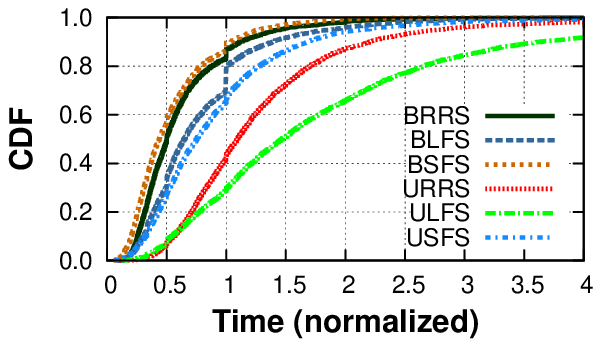}\label{subfig:phy2}}
	\subfigure[DBLP	($p=0.3$).]{\includegraphics[width= 0.32\textwidth]{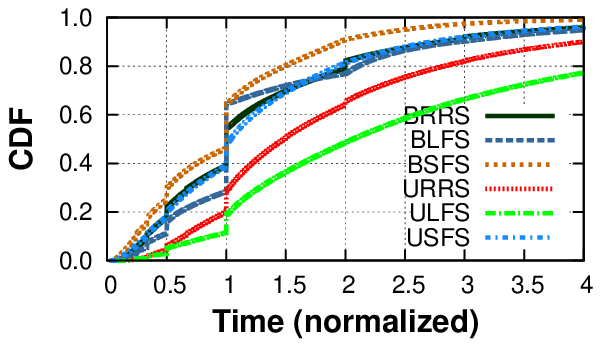}\label{subfig:phy3}}
	\subfigure[Epinion	($p=0.3$).]{\includegraphics[width= 0.32\textwidth]{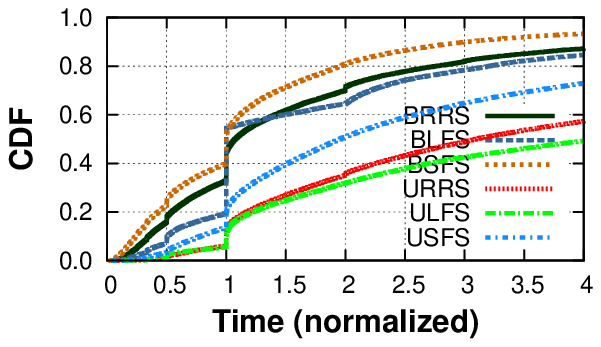}\label{subfig:phy3}}
	\subfigure[Wiki-vote	($p=0.3$).]{\includegraphics[width= 0.32\textwidth]{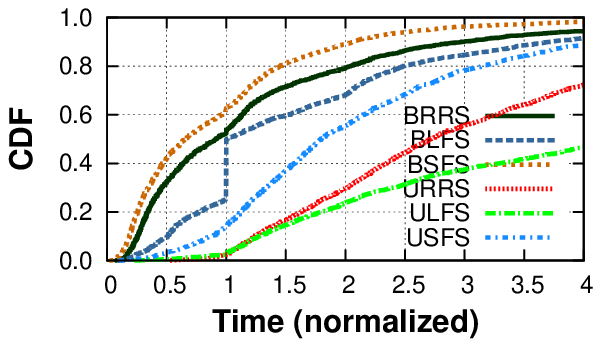}\label{subfig:phy3}}
\caption{The normalized time it takes to perform outsourced computations in \scloud, for variable task size.}\label{fig:var}
\end{center}
\end{figure*}

\subsubsection{Performance with different scheduling policies} Now, we turn our attention to measuring and understanding the impact of the different scheduling policies discussed in ~\textsection\ref{sec:taskpolicy} on the performance of \scloud. We consider the different datasets in Table~\ref{tab:datasets}, and use $p=0.1$ to $0.5$ with $0.2$ increments (the results are shown in Figure~\ref{fig:policy}). The observed consistent pattern in almost all figures in this experiment tells that shortest first policy always outperforms the round robin scheduling policy, whereas the round robin scheduling policy outperforms the longest first. This pattern is consistent regardless of $p$ and the outlier handling policy. 

The difference in the performance when using different policies can be as low as $2\%$ (when $p=0.1$ in physics co-authorship; shown in Figure~\ref{subfig:phy2}) and as high as $70\%$ (when using $p=0.5$ and outlier handling as in wiki-vote (figure~\ref{ppp10})). The patterns are made clearer in Figure~\ref{fig:policy} by observing combinations of parameters and policies.

We finally notice that, despite the difference in the performance of \scloud when using different policies, it still result in reasonable normalized finishing time to all users, suggesting its practicality against the measured metric with different parameters.

\subsubsection{Performance with Outliers Handling} Outliers, as defined in \textsection\ref{sec:outlier}, drag the performance of the entire system down. However, as pointed out earlier, handling outliers is quite simple in~\scloud~if accurate timing is used in the system. Here we consider the impact of the outlier handling policy explained in~\textsection\ref{sec:outlier} on the aggregate performance for the entire system. The impact of using the outlier handling policy can be also seen on Figure~\ref{fig:policy}, which is used for demonstrating the impact of using different scheduling policies as well. In this figure, we see that the simple handling policy we proposed improves the performance of the system greatly in all cases. 

More specifically, the improvement in the performance differs depending on other parameters, such as $p$, and the scheduling policy. As with the scheduling policy, the improvement can be as low as 2\% and as high as more than 60\%. When $p$ is large, the potential for improvement is high---see, for example, $p=5$ in Physics 2 (in Figure~\ref{fig:policy}) with the round robin scheduling policy where almost $65\%$ improvement is due to outlier handling when $x=1$. 

\subsubsection{Performance with Variable Task Size} In all of the above experiments, we considered computational tasks of fixed size; $1000$ of virtual time units in each of them. Whether the same pattern would be observed in tasks with variable size is unclear. Here we experimentally address this concern by using variable duty size that is uniformly distributed in the interval of $[500, 1500]$ time units. The results are shown in Figure~\ref{fig:var}. Comparing these results to the middle row of Figure~\ref{fig:policy} (for the fixed size tasks), we make two observations. (i) While the average task size in both scenarios is same, we observe that the performance with variable task size is worse. This performance is anticipated as our measure of performance is the time to finish that would be definitely increased as some tasks with longer time to finish are added. (ii) The same patterns advantaging a given scheduling policy on another are maintained as in earlier with fixed task length.

\subsubsection{Relationship Between Structure and Performance} It is worth noting that the performance of \scloud~is quite related to the underlying structure of the social graph. For example, sparse graphs such as co-authorship graphs---which are pointed out in~\cite{mohaisen11} to be slow mixing graphs---are the graphs with performance advantage in~\scloud. These graphs, in particular, are shown to possess a nice trust value that can be further utilized for \scloud. Furthermore, this trust value is unlikely to be found in online social networks which are prone to infiltration, making the case for trust-possessing graphs even stronger, as they achieve performance guarantees as well. This, indeed, is an interesting finding by itself, since it shows opposite outcomes to what is known in the literature on the usefulness of these graphs---see~\textsection\ref{sec:prelim} and more details, see~\cite{mohaisen11}. 

\subsection{Additional Features and Limitations}\label{sec:limits}
Our simulator of \scloud~omits a few details concerning the way a distributed system behaves in reality. In particular, our measurements do not report on or experiment with failure. However, our simulator is equipped with functionality for handling failure in the same way used for handling outliers (c.f.~\textsection\ref{sec:outlier}). Furthermore, our simulator considers a simplistic scenario of study by abstracting the hardware infrastructure, and does not consider additional resources consumed, such as memory and I/O resources. In the future, we will consider equipping our simulator with such functionalities and see how this affects the behavior and benefits of \scloud. 

One last concern related to our demonstration of our paradigm is that we do not consider the heterogeneity of resources, such as bandwidth and resources, in nodes acting as workers in the system. Furthermore, we did not consider how this affects the usability of our system and what decision choices this particular aspect of distributed computing systems would have on the utility of our paradigm. While this would be mainly a future work to consider (c.f.~\textsection\ref{sec:conclusion}), we expect that nodes would select workers among their social neighbors that have resources and link capacities exceeding a threshold, thus meeting an expected performance outcome. 
%Finally, for more experiments that combine several of the above scenarios, see the full technical report in~\cite{thetr}.

\section{Related Work}
\label{sec:related}

There have been many papers on the use of social networks for building communication and security systems, studying the performance of such designs on top of social networks, and analyzing the assumptions used in these designs as well. Below we highlight a few examples of these efforts and works.

Systems built on top of social networks include file sharing systems~\cite{isdal2010privacy}, anonymous communication systems~\cite{vasserman2009membership,Nagaraja07} Sybil defenses~\cite{DanezisM09,lesniewski2008sybil,YuGK07,YuKGF06}, referral and filtering systems~\cite{KautzSS97,resnick1994grouplens}, and live streaming~\cite{lin2009incentive}. Most of these applications weigh the trust in social graph, and an algorithmic property that makes the operation of these systems on top of social network effective. Another set of applications that exploit social networks' trust is routing~\cite{1544047,1288113,1281292,MartiGG04}---in several settings, where it has been shown that connectivity in social graphs can be of benefit in disconnected networks. Finally, assumptions of social network-based systems are explored recently, where Sybil defenses and their assumptions are studied in~\cite{mix-tr}, and trust is challenged in~\cite{mohaisen11}.

Perhaps the closest vein of related work in the literature to our work is on the use of social networks for building computing services. Until the time of writing this work, most of the prior research work has been solely focused on providing storage services, but not a platform of computations. Such storage services use slightly different economical model from \scloud's model, where payment per Megabyte per month rates are used as opposed to our eco-system. Examples of such efforts are reported by Sato~\cite{Sato09} and Tran et al.~\cite{tran2008friendstore}). Xu et al.~\cite{XuLP08} have further explored a first step in the direction of building cloud computing platforms on top of social networks where by considering the access control model in this domain with preferred access control guarantees. The results of this work can be used as a building block in our work to improve the quality of access control and authorization. 

With similar flavor of distributed computing services design, there has been prior works in literature on using volunteers' resources for computations exploiting locality of data~\cite{chandra,Weissman:2011:EED:1996014.1996019}, examination of programing paradigms, like MapReduce~\cite{Dean:2010aa} on such paradigm~\cite{Lin:2010:MMO:1851476.1851489,cardosa2011exploring}. Finally, our work shares several commonalities with the grid and volunteer computing systems~\cite{litzkow1988condor,Lin:2010:MMO:1851476.1851489,chandra,Weissman:2011:EED:1996014.1996019,anderson2002seti}, of which many aspects are explored in the literature. Trust of grid computing and volunteer-based systems is explored in~\cite{azzedin2002evolving,azzedin2002towards, song2005trusted,kamvar2003eigentrust,domingues2007sabotage}. Applications built on top of these systems, that would fit to our use model, are reported in~\cite{Weissman:2011:EED:1996014.1996019,cardosa2011exploring,wang2008scientific}, among others.

\section{Concluding Remarks}
\label{sec:discussion}
In this paper we have introduced the design of \scloud, a distributed computing service that recruits computing workers from friends in social networks and use such social networks that characterize trust relationships to bootstrap trust in the proposed computing service. We further advocated the case of such computing paradigm for the several advantages it provides. 

To demonstrate the potential of our proposed design, we used several real-world social graphs to bootstrap the proposed service and demonstrated that majority of nodes in most cases would benefit computationally from outsourcing their computations to such service. We considered several basic distributed system characteristics and features, such as outlier handling, scheduling decisions, and scheduler design, and show advantages in each of these features and options when used in our system. 

To the best of our knowledge, this is the first and only work in literature that bases such design of computing paradigm on volunteers recruited from social networks and tries to bring the trust factor from these networks and use it in such systems. This characteristic distances our work from the prior work in literature that uses volunteers' resources for computations~\cite{chandra,Weissman:2011:EED:1996014.1996019}.

Most important outcome of this study, along with the proposed design, is the relationship exposed between the social graphs and the behavior of the built computing service on top of them. In particular, we have shown that social graphs that possess strong trust characteristics as evidenced by face-to-face interaction~\cite{mohaisen11}, which are known in the literature for their poor characteristics prohibiting their use in applications (such as Sybil defenses~\cite{DanezisM09,YuGKX08,YuKGF06}), have a self-load-balancing characteristics when the number of outsourcers are relatively small (say $10$ to $20$ percent of the overall population on nodes in the computing services). That is, the time it takes to finish tasks originated by a given fraction of nodes in such graph, and for the majority of these nodes, ends in a relatively short time. 

On the other hand, such characteristics and advantages are maintained even when the number of outsourcers of computations is as high as $50\%$ of the nodes, contrary to the case of other graphs with dense structure and high connectivity known to be proper for the aforementioned applications. This last observation encourages us to investigate further scenarios of deployment of our design. We anticipate interesting findings based on the inherit structure of such deployment contexts---since such contexts may have different social structures that would affect the utility of the built computing overlay. %for \scloud~where other scheduling algorithms that weigh differential trust in these social networks are used. 

\newpage
\section{Future Work}\label{sec:future}
 
In the future we will look at two directions. In the first direction, we aim to complete the missing ingredient of the simulator and enrich it by further scenarios of deployment of our design, under failure, with different scheduling algorithms at both sides of the outsourcer and workers (in addition to those discussed in this work), and to consider other overhead characteristics that might not be in line with topological characteristics in the social graph. These characteristics may include the uptime, downtime, communication overhead, and I/O overhead consumption, among others. One interesting feature that we will consider is trust-based scheduling, benefiting from the prior work in~\cite{mohaisen11}.

In the second direction, we will turn our attention from the simulation settings to real-world deployment settings, thus addressing options discussed in~\textsection\ref{sec:limits}, and to implement a proof-of-concept application, among those discussed in~\textsection\ref{sec:usemodel}, by utilizing design options discussed in this paper. We anticipate a lot of hidden complexities in the design to arise, and significant findings to come out of the deployment that we will report on in the future work.

{%\small
\bibliographystyle{abbrv} 
\bibliography{refx}
}

\end{document}